\documentclass[aps,pra,reprint,superscriptaddress]{revtex4-1}

\usepackage[T1]{fontenc}
\usepackage[latin9]{inputenc}
\usepackage{textcomp}
\usepackage{amsmath}
\usepackage{amssymb}
\usepackage{graphicx}
\usepackage{color}
\usepackage{placeins}
\usepackage{hyperref}

\begin{document}

\title{Modeling charge relaxation in graphene quantum dots induced \\ by electron-phonon interaction}

\author{Sven Reichardt}
\affiliation{JARA-FIT and 2nd Institute of Physics, RWTH Aachen University, 52074 Aachen, Germany}
\affiliation{Physics and Materials Science Research Unit, Universit\'e du Luxembourg, 1511 Luxembourg, Luxembourg}
\
\author{Christoph Stampfer}
\affiliation{JARA-FIT and 2nd Institute of Physics, RWTH Aachen University, 52074 Aachen, Germany}
\affiliation{Peter Gr\"unberg Institute (PGI-9), Forschungszentrum J\"ulich, 52425 J\"ulich, Germany}

\begin{abstract}

We study and compare two analytic models of graphene quantum dots for calculating charge relaxation times due to electron-phonon interaction.
Recently, charge relaxation processes in graphene quantum dots have been probed experimentally and here we provide a theoretical estimate of relaxations times.
By comparing a model with pure edge confinement to a model with electrostatic confinement, we find that the latter features much larger relaxation times.
Interestingly, relaxation times in electrostatically defined quantum dots are predicted to exceed the experimentally observed lower bound of $\sim$100~ns.

\end{abstract}

\maketitle

\section{Introduction}

Graphene offers unique electronic properties~\cite{katsnelson2012} that make it a promising material for future nanoelectronics and quantum information technology.
Its low nuclear spin density (natural carbon consists of 99\% $^{12}$C, with no nuclear spin), leading to nearly negligible hyperfine interaction, and its weak spin-orbit interaction~\cite{huertas2006,min2006} make graphene also highly interesting for spintronics~\cite{han2014,roche2015} and spin-based applications.
In particular, the promise of long spin-relaxation times~\cite{recher2009} makes graphene an attractive candidate for hosting spin-based quantum bits~\cite{loss1998,trauzettel2007}. 
Motivated by this, graphene quantum dots (QDs) have been extensively investigated in recent years, both experimentally and theoretically.
While theoretical studies so far focused on, among others, level statistics~\cite{deraedt2008,libisch2009}, magnetic edge states~\cite{cheng2015}, and spin relaxation~\cite{struck2010}, the major experimental challenge lies in the confinement of charge carriers in graphene.
In fact, the gapless electronic band structure and Klein tunneling effects~\cite{katsnelson2006,beenakker2008} make the purely electrostatic confinement of charge carriers in graphene a difficult task.
Instead, the most viable way to confine charge carriers in graphene nowadays is based on etching nanostructures into graphene flakes~\cite{stampfer2011}.
So far, Coulomb blockade~\cite{stampfer2008,ponomarenko2008,liu2009,puddy2013}, excited states~\cite{schnez2009,volk2011,guttinger2009}, charge sensing~\cite{wang2010,guttinger2011}, and spin-filling sequences~\cite{guttinger2010} have been studied in detail in etched graphene quantum dot devices. 
More recently, also charge pumps~\cite{connolly2013} and charge relaxation times of excited states~\cite{volk2013} in graphene quantum dots have been investigated experimentally.
The extracted lifetimes are a factor of 5 -- 10 larger, with an extracted lower bound of around 60 -- 100~ns compared to III/V quantum dots~\cite{hanson2007,fujisawa2002,jang2008}.
It has been suggested~\cite{volk2013} that the increased lifetimes of excited states in graphene can be attributed to weak electron-phonon interaction due to the absence of piezo-electric phonon modes in graphene.
However, a detailed theoretical estimate of electron-phonon interaction-induced charge relaxation in graphene has been missing so far.

Here we present and discuss analytical calculations of charge relaxation times in graphene quantum dots.
In particular, we compare two simple models of QDs that have recently been discussed in the literature~\cite{recher2009,schnez2008,struck2010}.
These studies focused on spin-relaxation processes and assumed a ratio of QD radius-to-phonon wavelength much smaller than the one reported in Ref.~\onlinecite{volk2013}.
These results are thus not applicable to the charge relaxation processes in large QDs.
In this paper, after briefly reviewing both models, we derive analytical expressions for charge relaxation times due to electron-phonon interaction, going beyond the previously used dipole approximation for the electron-phonon coupling.
We find that the quantitative results significantly differ between the two models, which can be traced back to the different way in which confinement is handled.
In particular, we find that, in a model where electrons are confined electrostatically, charge relaxation times are much larger than in other models, suggesting electrostatic confinement as a preferred way to define quantum dots in graphene.

\section{Graphene Quantum Dot Models}

We start by briefly reviewing the two models used for the calculation of charge relaxation times.
In both models the quantum dot is taken to be circular, while the method of confinement differs between the two (see Figures~1a and 1b).

\begin{figure}[t]
\includegraphics[width=1.0\linewidth]{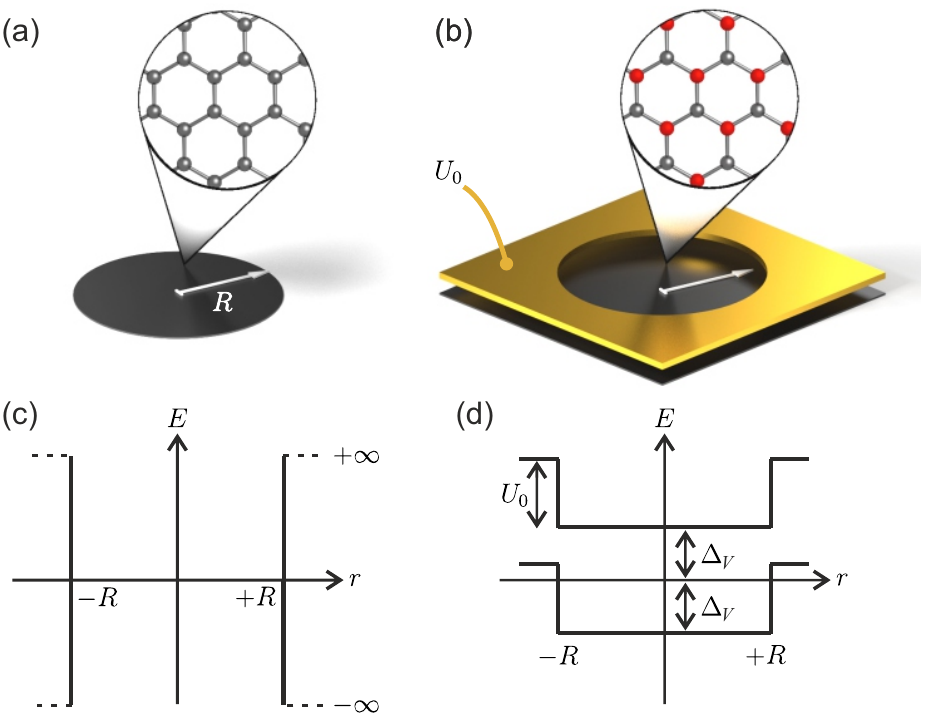}
\caption{(a,b) Schematic illustrations of the two different graphene quantum dot models describing a circular dot with radius $R$: 
(a) Infinite mass boundary model and (b) Electrostatic confinement model.
(c,d) The corresponding total potential for both models. The quantities appearing in the labels are defined in the text.
\label{fig:potential}
}
\end{figure}

In the first model, which we will refer to as the ``infinite mass boundary'' (IM) model, the electron behaves like a free electron inside the dot, while being confined by an infinitely large mass-like potential~\cite{schnez2008} at the boundary that might be realized by etching the dot.
The second model, which will be referred to as the ``electrostatic confinement'' (EC) model, uses a finite electrostatic potential to achieve confinement~\cite{recher2009}.
To prevent Klein tunneling, we account for a finite mass term that might be induced by an underlying substrate.
Within the nearest-neighbor tight-binding scheme in the low-energy approximation, both models can be described by a similar Hamiltonian in the valley isotropic form~\cite{schnez2008,recher2009}:
\begin{equation}
\mathbf{H}^{\tau_{\mathrm{v}}} = v_{\mathrm{F}} \vec{p} \cdot \vec{\sigma} + \tau_{\mathrm{v}} V(r) \sigma_z + U(r).
\label{eq:Hamiltonian}
\end{equation}
Here, $v_{\mathrm{F}} \approx 10^6$~m/s is the Fermi velocity, $\vec{p}$ is the momentum operator, $\vec{\sigma} = (\sigma_x,\sigma_y)^{\mathrm{T}}$ is a two-dimensional vector of Pauli matrices acting in sublattice space, and $\tau_{\mathrm{v}} = \pm 1$ distinguishes between the $K$ and $K'$ valleys, respectively.
$V(r)$ specifies the mass term and only depends on the radial coordinate for a circularly symmetric quantum dot.
For the IM~model, $V(r)$ is given by $V_{\mathrm{IM}}(r) = 0$ for $r < R$ and $V_{\mathrm{IM}}(r) = \infty$ for $r > R$, while we use $V_{\mathrm{EC}}(r) = \Delta_V = \mathrm{const}$ for $V(r)$ in the EC~model.
Finally, $U(r)$ denotes the electrostatic confinement potential, which is zero in the IM~model.
For the EC~model, we use $U_{\mathrm{EC}}(r) = 0$ for $r < R$ and $U_{\mathrm{EC}}(r) = U_0 = \mathrm{const}$ for $r > R$.
The different potential landscapes are illustrated in Figures~1c and 1d.

The Hamiltonian given in Eq.~(\ref{eq:Hamiltonian}) commutes with the operator of total angular momentum $\mathbf{J}_z = L_z + (\hbar/2)\sigma_z$, where $L_z = (\vec{r} \times \vec{p})_z$ denotes the $z$~component of the orbital angular momentum operator and the second term corresponds to the contribution of the pseudospin (or sublattice spin) in graphene.
The eigenfunctions of the Hamiltonian can thus be chosen to also be eigenfunctions of $\mathbf{J}_z$ and hence we make the ansatz
\begin{equation}
\psi(r,\varphi) = \left(
\begin{array}{c}
\psi_{m}^{(1)}(r,\varphi) \\
\psi_{m}^{(2)}(r,\varphi)
\end{array}
\right) = \left(
\begin{array}{c}
\chi_{m}^{(1)}(r) \\
\chi_{m}^{(2)}(r) \mathrm{e}^{i \varphi}
\end{array}
\right) \mathrm{e}^{i m \varphi},
\end{equation}
with $m \in \mathbb{Z}$ being the angular momentum quantum number.
The Dirac equation $(\mathbf{H}^{\tau_{\mathrm{v}}}-E)\psi(r, \varphi) = 0$ has been solved for both models in previous works~\cite{schnez2008,recher2009}.
In both cases, the above ansatz results in a simple, Bessel-type differential equation for $\chi_{m}^{(1,2)}(r)$.
The boundary conditions, however, are different in both models.

In the IM~model, the wave functions are restricted by the condition that the radial component of the probability current is zero, so that the electron remains within the dot at all times.
This results in the condition $\psi_{m}^{\tau_{\mathrm{v}},(2)}(R,\varphi) / \psi_{m}^{\tau_{\mathrm{v}},(1)}(R,\varphi) = \tau_{\mathrm{v}} i\mathrm{e}^{i \varphi}$~\cite{berry1987}.
One then finds that the possible wave functions in the IM~model are given by
\begin{equation}
\psi_{m,n}^{\tau_{\mathrm{v}}}(r,\varphi) =
\begin{cases}
\alpha \left(
\begin{array}{c}
J_m(k r) \\
i J_{m+1}(k r) \mathrm{e}^{i\varphi}
\end{array}
\right) \mathrm{e}^{i m \varphi}, & r < R \\
0, & r \geq R,
\end{cases}
\end{equation}
where $k = E / (\hbar v_{\mathrm{F}})$, $J_m$ denotes the Bessel function of the first kind of order $m$, and $\alpha > 0$ is a normalization constant.
The boundary condition then translates to the transcendental equation
\begin{equation}
J_{m+1}(k R) = \tau_{\mathrm{v}} J_m(k R),
\label{eq:IM-boundary}
\end{equation}
which has solutions for discrete values of $k$ that we will label by an additional, positive integer quantum number $n$, sorted by increasing energy.
Due to the oscillatory nature of Bessel functions, this equation has an infinite number of solutions.
Note also that the wave function is discontinuous at $r = R$, as the mass term features an infinitely high step there, so that the first derivative of the wave function has a $\delta$-function-like behavior at $r = R$.
This leads to a discontinuity in the wave function itself, in contrast to the behavior of the wave function for a non-relativistic particle, which is continuous even for potentials containing a $\delta$-like singularity since the equation of motion features the second spatial derivative of the wave function.
This behavior is illustrated in Figure~2a, where one example of such a wave function is plotted.

\begin{figure}[t]
\includegraphics[width=1.0\linewidth]{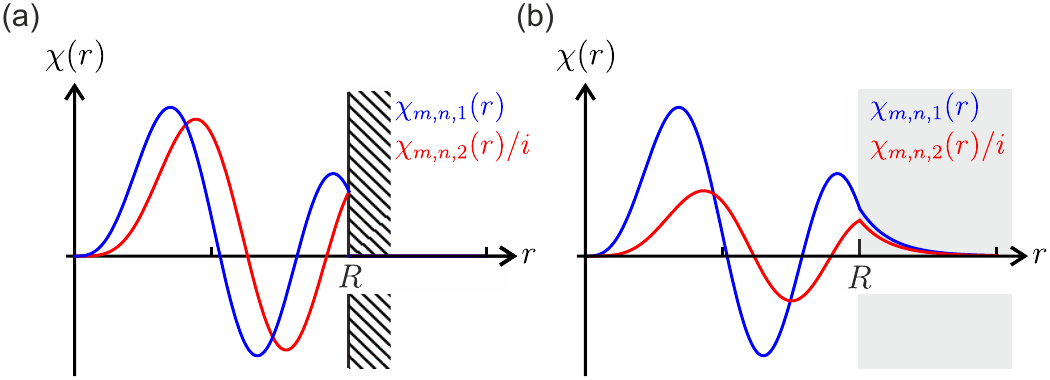}
\caption{
Examples of typical wave functions in the two models.
(a) Infinite mass boundary model.
(b) Electrostatic confinement model.
The quantum numbers belonging to the shown wave functions are $\tau_{\mathrm{v}} = +1$ ($K$~valley), $m = 3$ and $n = 3$ for both cases.
In case of the EC~model, the parameters were set to $\Delta_V = U_0 = 10\hbar v_{\mathrm{F}} / R \approx 120$~meV, following Ref.~\onlinecite{struck2010}.
\label{fig:wave-function}
}
\end{figure}

By contrast, all potentials in the EC~model only possess finite steps and hence the allowed wave functions must be continuous at $r = R$.
The general solution of the Dirac equation in this case is given by
\begin{widetext}
\begin{equation}
\psi_{m,n}^{\tau_{\mathrm{v}}}(r,\varphi) =
\begin{cases}
\alpha \left(
\begin{array}{c}
J_m\left(\sqrt{k_<^2} r\right) \\
i \sqrt{\frac{\kappa_< - \lambda_{\tau_{\mathrm{v}}}}{\kappa_< + \lambda_{\tau_{\mathrm{v}}}}} J_{m+1}\left(\sqrt{k_<^2} r\right) \mathrm{e}^{i \varphi}
\end{array}
\right) \mathrm{e}^{i m \varphi}, & r < R \\
\alpha \eta \left(
\begin{array}{c}
K_m\left(\sqrt{-k_>^2} r\right) \\
i \sqrt{\frac{\lambda_{\tau_{\mathrm{v}}} - \kappa_>}{\lambda_{\tau_{\mathrm{v}}} + \kappa_>}} K_{m+1}\left(\sqrt{-k_>^2} r\right) \mathrm{e}^{i \varphi}
\end{array}
\right) \mathrm{e}^{i m \varphi}, & r \geq R.
\end{cases}
\end{equation}
\end{widetext}
Here we introduced $\lambda_{\tau_{\mathrm{v}}} = \tau_{\mathrm{v}}\Delta_V / (\hbar v_{\mathrm{F}})$, $\kappa_< = E / (\hbar v_{\mathrm{F}})$, $\kappa_> = (E-U_0) / (\hbar v_{\mathrm{F}})$, and $k_{<,>}^2 = \kappa_{<,>}^2 -  \lambda_{\tau_{\mathrm{v}}}^2$.
The function $K_m$ is the modified Bessel function of the first kind of order $m$, while the constant $\alpha > 0$ is again a normalization constant.
Finally, the constant $\eta$ is needed to ensure that both components of the spinor are continuous at $r = R$.
Explicitly, it is given by $\eta = J_m(\sqrt{k_<^2} R)/K_m(\sqrt{-k_>^2} R)$.
By equating the lower components of the wave functions at $r = R$, one obtains an equation that determines the possible energy values:
\begin{equation}
\begin{split}
& \sqrt{\frac{\kappa_< - \lambda_{\tau_{\mathrm{v}}}}{\kappa_< + \lambda_{\tau_{\mathrm{v}}}}} J_{m+1}\left(\sqrt{k_<^2} R\right) K_m\left(\sqrt{-k_>^2} R\right) \\
= & \sqrt{\frac{\lambda_{\tau_{\mathrm{v}}} - \kappa_>}{\lambda_{\tau_{\mathrm{v}}} + \kappa_>}} J_m\left(\sqrt{k_<^2} R\right) K_{m+1}\left(\sqrt{-k_>^2} R\right).
\end{split}
\end{equation}
In contrast to the characteristic equation for the IM~model [Eq.~(\ref{eq:IM-boundary})], the above equation only has a finite number of roots, as the modified Bessel functions are exponentially damped.
An example of a typical wave function is shown in Figure~2b.
Compared to the wave functions in the IM~model, the EC~wave functions feature a leaking tail due to the finite confinement potential.
Having reviewed the two models and the resulting wave functions, we can now move on to calculate charge relaxation times due to electron-phonon interaction.

\section{Electron-Phonon Interaction and Charge Relaxation Times}

Electron-phonon interaction plays an important role in many graphene experiments.
For example, electron-phonon scattering has been shown to limit the charge-carrier mobility in graphene samples~\cite{hwang2008} and has also been argued to be the main mechanism for charge relaxation in QDs~\cite{volk2013}.
To gain an estimate of the charge relaxation times in graphene quantum dots, we thus focus on relaxation processes due to electron-phonon interaction.
Theoretically, electron-phonon interaction in graphene has already been studied in the literature~\cite{suzuura2002,ando2005,manes2007,mariani2008} in depth.
We use a perturbative approach and look only at the dominant, lowest-order, i.e.. one-phonon, processes.
Since we want to provide an estimate for charge relaxation times observed in an experiment at low temperature~\cite{volk2013}, we can neglect phonon absorption.
Thus, the only process we consider here is the decay of an excited QD state to a lower state by emission of one phonon.
Furthermore, the energy differences between the electronic states observed in experiment are rather low ($\Delta_E \lesssim 10$~meV~\cite{volk2013}) and thus we only take into account low-energy, acoustic phonons.
Moreover, since out-of-plane lattice vibrations couple only via two-phonon processes~\cite{manes2007}, we can further restrict ourselves to in-plane phonon modes.
The latter couple to the electronic system via two mechanisms: (i) a deformation potential (DEF) caused by a change of the area of the unit cell, and (ii) a change of the carbon-atom bond length (BLC) resulting in a modified hopping parameter.
Since we are interested in one-phonon processes involving acoustic, low-energy (i.e., low-momentum) phonons, inter-valley scattering is not possible and we can thus focus on only one valley, choosing the one at $K$.
Following the approach used in Ref.~\onlinecite{mariani2008}, we find that to lowest order in the lattice deformation field, the electron-phonon interaction Hamiltonian can be written as
\begin{equation}
\begin{split}
H_{\mathrm{el-ph}} = \sum_{\substack{\mathrm{a} = \mathrm{DEF},\\\mathrm{BLC}}} g_{\mathrm{a}} \sum_{\substack{m',n'\\ m,n}} & \sum_{\vec{q},\mu} f_{\mu}(q) (b_{\vec{q},\mu} + b_{-\vec{q},\mu}^{\dagger}) \\
& \times c_{m',n'}^{\dagger} c_{m,n} \mathcal{M}_{m',n',m,n}^{\mathrm{a},\mu}(\vec{q}).
\end{split}
\end{equation}
Here, $g_{\mathrm{a}}$ is the electron-phonon coupling constant related to mechanism $\mathrm{a} = \mathrm{DEF,BLC}$.
Numerically, we use $g_{\mathrm{DEF}} = 30$~eV and $g_{\mathrm{BLC}} = 1.4$~eV~\cite{suzuura2002,struck2010}.
We also defined $f_{\mu}(q) = \sqrt{\hbar / (A \rho \omega_{\mu}(q))} q$, wherein $A$ is the area of the graphene sheet, $\rho$ is its mass density, and $\omega_{\mu}(q)$ is the frequency of an in-plane phonon with wave vector $\vec{q} = q (\cos \varphi_{\vec{q}},\sin \varphi_{\vec{q}})^{\mathrm{T}}$ and longitudinal or transverse polarization $\mu = \mathrm{l}, \mathrm{t}$.
The operators $b_{\vec{q},\mu}$($b_{\vec{q},\mu}^{\dagger}$) and $c_{m,n}$ ($c_{m,n}^{\dagger}$) are the phononic and electronic annihilation (creation) operators, which destroy (create) a phonon or electron with the indicated quantum numbers, respectively.
Finally, $\mathcal{M}_{m',n',m,n}^{\mathrm{a},\mu}(\vec{q})$ denotes the matrix element for the involved states.
It can be written in the form
\begin{equation}
\begin{split}
\mathcal{M}_{m',n',m,n}^{\mathrm{a},\mu}(\vec{q}) = \int \mathrm{d}^2 r \, & \psi_{m',n'}^*(\vec{r}) \left(
\begin{array}{cc}
\alpha_{\mathrm{a},\mu} & \beta_{\mathrm{a},\mu} \\
\beta_{\mathrm{a},\mu}^* & \alpha_{\mathrm{a},\mu}
\end{array}
\right) \psi_{m,n}(\vec{r}) \\
& \times\mathrm{e}^{i \vec{q} \cdot\ \vec{r}},
\end{split}
\label{eq:matrix-element}
\end{equation}
where the constants $\alpha_{\mathrm{a},\mu}$ and $\beta_{\mathrm{a},\mu}$ are given by $\alpha_{\mathrm{a} = \mathrm{DEF}, \mu = \mathrm{l}} = -1$, $\beta_{\mathrm{BLC},\mathrm{l}} = \exp(2 i \varphi_{\vec{q}})$, $\beta_{\mathrm{BLC},\mathrm{t}} = i \exp(2 i \varphi_{\vec{q}})$ and are zero for all other cases~\cite{suzuura2002}.

To facilitate the evaluation of these matrix elements, the exponential in Eq.~(\ref{eq:matrix-element}) has so far been approximated by the dipole approximation in the recent literature~\cite{struck2010}.
This is possible if the wavelength of the phonon is much larger than the QD radius.
In the experiment which we want to compare our results to, however, the QD radius is given by $R = 55$~nm\cite{volk2013}, which is larger than the typical wavelength of the involved phonons ($\lambda \approx 5-10$~nm, for a phonon of energy $\hbar \omega = \Delta_E \lesssim 10$~meV).
Thus, the dipole approximation for $\exp(i \vec{q} \cdot \vec{r})$ is no longer applicable.
Instead, one can make use of the Jacobi-Anger identity $\exp( i \vec{q} \cdot \vec{r}) = \exp(i q r \cos(\varphi-\varphi_{\vec{q}})) = \sum_{l = -\infty}^{+\infty} i^l J_l(q r) \mathrm{e}^{i l (\varphi-\varphi_{\vec{q}})}$ to expand the exponential of Eq.~(\ref{eq:matrix-element}) into a Fourier series.
After evaluating the angular part of the integral, the matrix elements reduce to:
\begin{widetext}
\begin{equation}
\mathcal{M}_{m',n',m,n}^{\mathrm{a},\mu}(\vec{q}) = -2 \pi i^{m'-m} \mathrm{e}^{-i (m'-m) \varphi_{\vec{q}}} \times
\begin{cases}
M_0^{(1,1)}(q) + M_0^{(2,2)}(q), & \mathrm{a} = \mathrm{DEF}, \mu = \mathrm{l} \\
0, & \mathrm{a} = \mathrm{DEF}, \mu = \mathrm{t} \\
i M_{-1}^{(1,2)}(q) \mathrm{e}^{+3 i \varphi_{\vec{q}}} - i M_{+1}^{(2,1)}(q)\mathrm{e}^{-3 i \varphi_{\vec{q}}}, & \mathrm{a} = \mathrm{BLC,\mu = \mathrm{l}} \\
-M_{-1}^{(1,2)}(q) \mathrm{e}^{+3 i \varphi_{\vec{q}}} - M_{+1}^{(2,1)}(q)\mathrm{e}^{-3 i \varphi_{\vec{q}}}, & \mathrm{a} = \mathrm{BLC},\mu = \mathrm{t},
\end{cases}
\end{equation}
\end{widetext}
where we introduced $M_{\zeta}^{(i,j)}(q) = \int_0^{\infty} \mathrm{d} r \, r \chi_{m',n'}^{(i),*}(r) \chi_{m,n}^{(j)}(r) J_{m'-m+\zeta}(q r)$ as a short-hand notation that hides the dependence of the matrix element on $m$, $n$, $m'$, and $n'$, but avoids a cluttering of the final expressions due to too many indices.

Finally, we can make use of Fermi's golden rule,
\begin{widetext}
\begin{equation}
\tau^{-1}((m,n) \to (m',n')) = \frac{2 \pi}{\hbar} \sum_{\vec{q},\mu} | \langle m',n';\mu,\vec{q}| H_{\mathrm{el-ph,ac}} |m,n\rangle |^{2} \times \delta(E_{m,n} - E_{m'n'} - \hbar \omega_{\mu}(q)),
\label{eq:relax-time}
\end{equation}
\end{widetext}
to get an analytic expression for the charge relaxation time $\tau$.
For low-energy acoustic phonons, we use a linear approximation of the phonon dispersion, $\omega_{\mu}(q) = v_{\mu} q$, where $v_{\mathrm{l}} \approx 19.5 \times 10^3$~m/s and $v_{\mathrm{t}} \approx 12.2 \times 10^3$~m/s is the speed of sound for longitudinal and transverse phonons, respectively \cite{struck2010}.
After approximating the sum over $\vec{q}$ in Eq.~(\ref{eq:relax-time}) with an integral $\sum_{\vec{q}} \to \frac{A}{(2 \pi)^2} \int \mathrm{d}^2 q$, any interference terms in the squared matrix elements disappear as they contain oscillating exponentials of $\varphi_{\vec{q}}$.
The final result for the relaxation times reduces to a sum of three contributions $\tau^{-1}((m,n) \to (m',n')) = \tau^{-1}|_{\mathrm{DEF,l}} + \tau^{-1}|_{\mathrm{BLC,l}} + \tau^{-1}|_{\mathrm{BLC,t}}$, given by:
\begin{widetext} 
\begin{equation}
\tau^{-1}|_{\mathrm{a},\mu} =
\begin{cases}
\theta(\Delta_E) \frac{4 \pi^2 \Delta_E^2}{\hbar^3 v_{\mathrm{l}}^4 \rho} \left|M_0^{(1,1)}\left(\frac{\Delta_E}{\hbar v_{\mathrm{l}}}\right) + M_0^{(2,2)}\left(\frac{\Delta_E}{\hbar v_{\mathrm{l}}}\right)\right|^2, & \mathrm{a} = \mathrm{DEF}, \mu = \mathrm{l} \\
\theta(\Delta_E) \frac{4 \pi^2 \Delta_E^2}{\hbar^3 v_{\mathrm{l}}^4 \rho} \left[\left|M_{-1}^{(1,2)}\left(\frac{\Delta_E}{\hbar v_{\mathrm{l}}}\right)\right|^2 + \left|M_{+1}^{(2,1)}\left(\frac{\Delta_E}{\hbar v_{\mathrm{l}}}\right)\right|^2\right], & \mathrm{a} = \mathrm{BLC},\mu = \mathrm{l} \\
\theta(\Delta_E) \frac{4 \pi^2 \Delta_E^2}{\hbar^3 v_{\mathrm{t}}^4 \rho} \left[\left|M_{-1}^{(1,2)}\left(\frac{\Delta_E}{\hbar v_{\mathrm{t}}}\right)\right|^2 + \left|M_{+1}^{(2,1)}\left(\frac{\Delta_E}{\hbar v_{\mathrm{t}}}\right)\right|^2\right], & \mathrm{a} = \mathrm{BLC},\mu = \mathrm{t},
\end{cases}
\end{equation}
\end{widetext}
where $\Delta_E = E_{m,n} - E_{m'n'}$ denotes the energy difference between the initial and final electron state and $\theta(\Delta_E)$ is the step function

\section{Numerical Results and Discussion}

\begin{figure*}[!]
\includegraphics[width=1.0\linewidth]{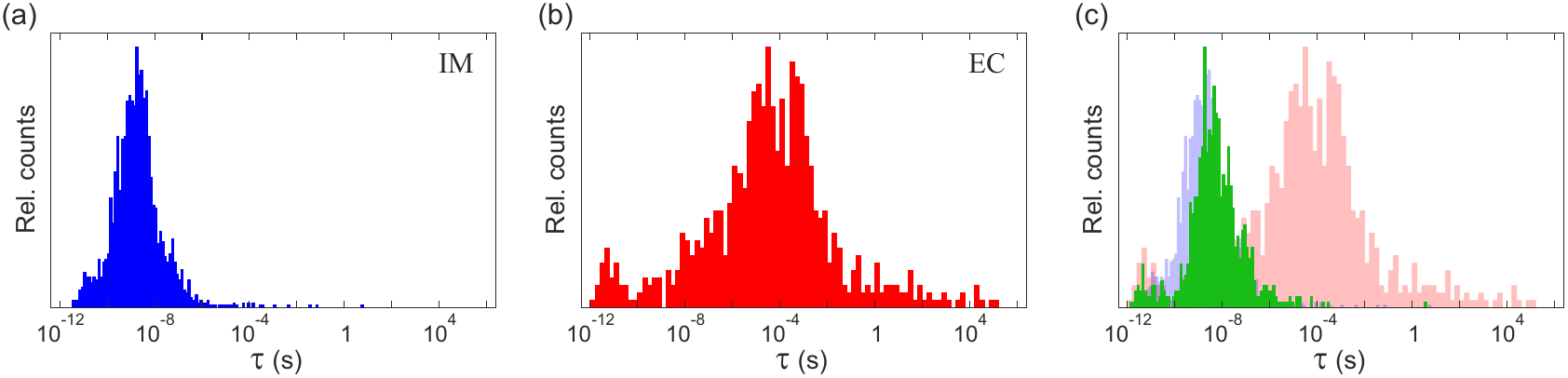}
\caption{
Numerical results for the charge relaxation times for the finite sets of states described in the text.
(a) Infinite mass boundary model.
(b) Electrostatic confinement model.
(c) Electrostatic confinement model, with the wave function artificially set to zero outside the quantum dot (green histogram).
For comparison, the distributions (histograms) shown in panels (a) and (b) are also shown.
\label{fig:Relax-Histo}
}
\end{figure*}

To obtain an estimate of the order of magnitude of charge relaxation times in graphene quantum dots, we apply the expressions obtained in the previous section to the electronic states of the two quantum dot models.
For both models, we use a dot radius of $R = 55$~nm and, in the case of the electrostatic confinement model, we follow Ref.~\onlinecite{struck2010} and use $\Delta_V = U_0 = 10\hbar v_{\mathrm{F}} / R \approx 120$~meV.
To obtain numerical results for the relaxation times, we need to limit the number of states to a finite number.
For the EC~model, the number of states is naturally limited by the finite size of the well and hence we include all of them in the numerical analysis.
For the IM~model, which features an infinitely high mass barrier, no such natural limitation exists and the number of states has to be chosen manually.
To allow for a comparison with experiment~\cite{volk2013}, we focus on low-lying excited states in the conduction band.
Furthermore, in order to facilitate the comparison between the two models, we choose an energy cutoff $E^{\mathrm{IM}}_{\mathrm{max}}$ for the states in the IM~model.
This energy cutoff can be chosen to either match the maximum possible energy in the EC~model or to result in roughly the same number of included states as in the EC~model.
However, the former possibility, i.e., $E^{\mathrm{IM}}_{\mathrm{max}} = \Delta_V \approx 120$~meV, results in an unfeasibly  small number of states due to a comparatively large level spacing in an infinitely deep well.
We therefore limit the energy of the states such that the number of states in the conduction band is roughly the same in both models, which results in a chosen maximum energy of $E^{\mathrm{IM}}_{\mathrm{max}} = 250$~meV.
The highest possible angular momentum quantum numbers turn out to be $m_{\mathrm{max,IM}} = 18$ and $m_{\mathrm{max,EC}} = 13$.
For each model, we compute the relaxation times between all possible pairs of states in the conduction band and collect the results in a histogram shown in Figures~3a and 3b.
In both models, the calculated times range over several orders of magnitude.
However, the range of $\tau$ values that occur most often differs significantly between the two models.
In the IM~model, one mostly finds relaxation times between 1~\textmu s and 0.1~ns, whereas the EC~model results in a wide spread of times.
In the latter, the majority of times fall into the interval between 10~ms and 10~ns, thus predicting the existence of long-lived excited states.
This behavior has its origin in the finite probability to find the electron outside the dot, since it can be traced back to the presence of a leaking tail of the wave function.
To demonstrate this, we set the leaking tail of the wave function in the EC~model artificially to zero, renormalize the wave function, and recalculate the relaxation times.
In Figure~3c, we show the resulting distribution of times as a green histogram.
When compared to the two distributions of the IM and EC~model (half-transparent blue and red histograms, respectively), we find that the resulting distribution is much closer to the one found in the IM~model, where no leaking tail is present.
Further analysis of the numerical results shows that the contribution of the region $r \geq R$ is roughly of the same order of magnitude as the one of the region $r \leq R$.
However, the signs of the two contributions are opposite, and hence the average electron-phonon matrix element in the EC~model is rather small.
This leads to the (on average) small decay widths and correspondingly large life times in the EC~model.
Unfortunately, the physical origin of this effect is not easy to pinpoint, especially since the sign and magnitude of the contributions of the two different regions to the matrix element strongly depend on the concrete quantum numbers of the involved states.

While the total relaxation times differ significantly between the two models, the relative contributions of the two different electron-phonon coupling mechanisms are roughly the same in both cases.
If one calculates the individual contributions of both the deformation potential and the bond length change mechanism to the decay width, one finds that the former is on average a factor of $10^6$ larger than the latter.
This factor is three orders of magnitudes larger than what would be expected from the ratio of the squared coupling constants alone and indicates that the form of the electronic wave functions and their interplay with the phonon play a significant role.
In other words, the change of the on-site energies (which corresponds to the deformation potential mechanism) is much more important than the change of the hopping parameters for the localized states of the quantum dot.

\begin{figure*}[!]
\includegraphics[width=1.0\linewidth]{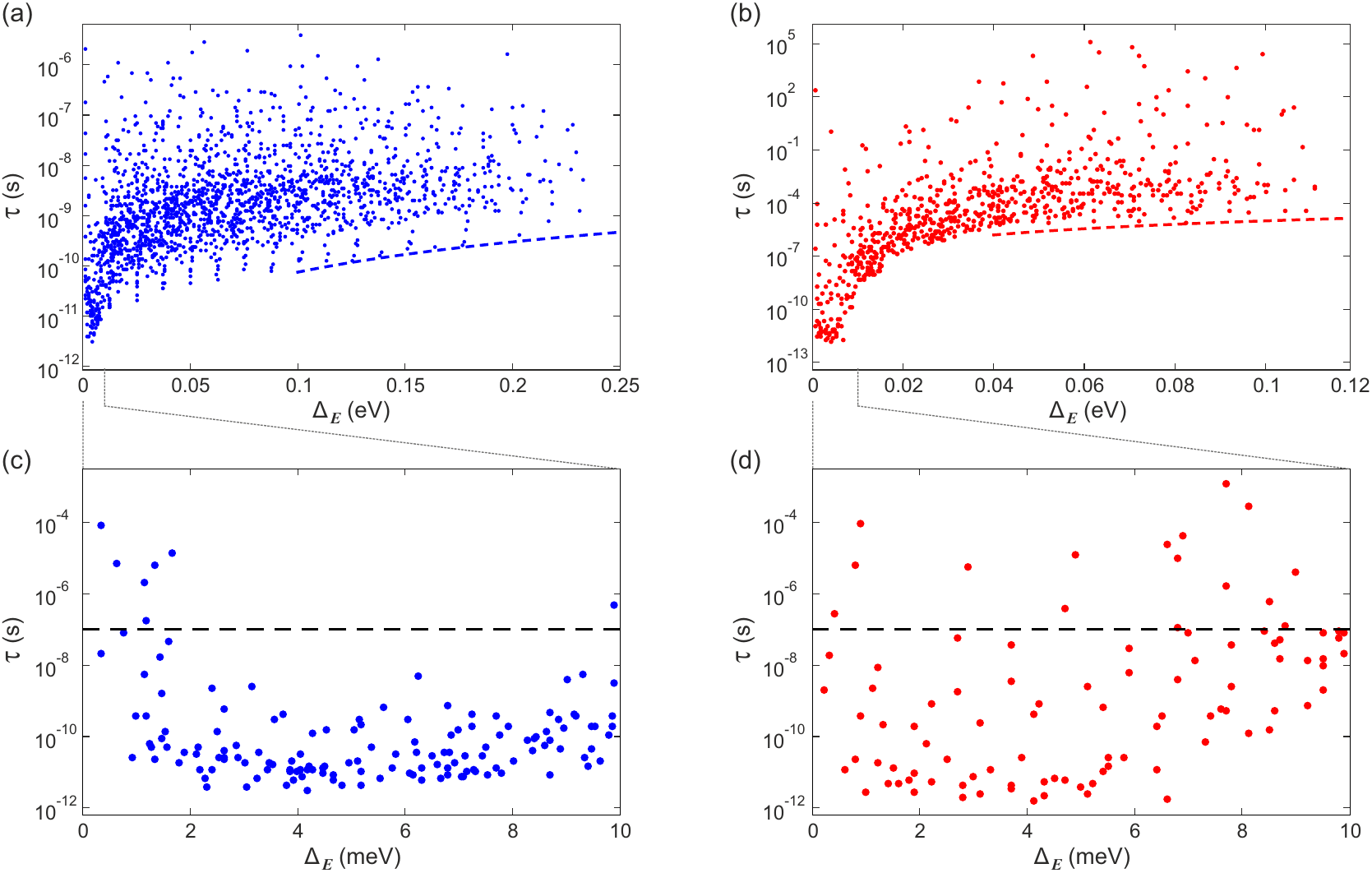}
\caption{
Level spacing versus charge relaxation times in the two quantum dot models.
(a) Infinite mass boundary model.
(b) Electrostatic confinement model.
The dashed lines serve as a guide to the eye for the asymptotic $\tau \sim \Delta_E^2$ behavior.
(c) and (d) Zoom-in into the region $\Delta_E < 10$~meV of panels (a) and (b), respectively.
The dashed line represents the experimentally observed lower bound as discussed in Ref.~\onlinecite{volk2013}.
\label{fig:Delta-vs-Relax}
}
\end{figure*}

For further analysis and comparison to experiment, we plot the relaxation times against the level spacing as shown in Figures~4a and 4b.
Here, one can observe the general trend that the calculated relaxation times as a whole grow monotonically with the level spacing.
For large values of the level spacing $\Delta_E$, one finds a lower bound obeying a $\tau \sim \Delta_E^2$ scaling law (compare dashed lines in Figures~4a and 4b).
This behavior can be directly extracted from the analytical calculation, since for large $\Delta_E$, the integrals $M_{\zeta}^{(i,j)}(\Delta_E / \hbar v_{\mathrm{F}})$ scale like $\Delta_E^{-2}$ and hence the decay rate $\tau^{-1}$ scales like $\Delta_E^2 \times [\Delta_E^{-2}]^2 = \Delta_E^{-2}$.
Also notable is the fact that even for small level spacings ($ \Delta_E < 10$~meV), i.e., in the experimentally relevant region~\cite{volk2013}, the two models lead to rather different predictions as seen more clearly in the zoom-in shown in Figures~4c and 4d.
In this region, only the EC~model consistently features excited states with life times above the lower bound observed in experiment (black, dashed line in Figures~4c and 4d).

At this point it is important to note that the experimental data from Ref.~\onlinecite{volk2013} were obtained on etched graphene nanostructures.
The termination of the nanostructure corresponds to the appearance of an infinitely high mass term and thus, out of the two investigated models, the IM~model is expected to correspond best to the experimental situation.
However, the fact that the average relaxation times in the IM~model are much lower than the lower bound observed in the cited experimental work seems to suggest that this simple model does not capture all of the relevant physics, even though it should be noted that it does feature transitions with relaxation times above the lower bound in the region $\Delta_E < 2$~meV, which partly contains the excited states seen in experiment (with energies in the range 1.7 -- 2.5~meV~\cite{volk2013}).
It is also possible that near the edges of the graphene nanostructure the sublattice symmetry is broken (e.g., due to the badly controlled etching procedure).
This would lead to a finite mass term toward the boundary of the dot, which would in turn lead to a leaking tail of the wave function.
As observed in the EC~model, this leaking tail could then be responsible for the large increase of the relaxation times.
Independently of this, the larger relaxation times present in the EC~model suggest that the presence of a leaking tail is beneficial for the longevity of the excited states of the quantum dot and hence electrostatic confinement might be the better way to define quantum dots with larger charge relaxation times.

In summary, we compared two models of graphene quantum dots and studied charge relaxation times due to emission of acoustic phonons.
We found that the two models lead to significantly different predictions that could be traced back to the different boundary conditions and the corresponding existence of a leaking tail of the wave functions.
Furthermore, we demonstrated that the relaxation times in the electrostatic confinement model are much larger than the times found in the infinite mass boundary model, exceeding the experimentally established lower boundary of 60 -- 100~ns.
Thus, for the realization of graphene quantum dots with long-living excited states, the confinement with electrostatic potentials as featured in the model, is to be preferred.

\section*{Acknowledgments}

The authors would like to thank G. Burkard, F. Haupt, and S.V. Rotkin for useful discussions.
Financial support by the Deutsche Forschungsgesellschaft (Grant No. SPP-1459) and the European Research Council (Grant No. GA-Nr. 280140) is gratefully acknowledged.
S. R. acknowledges financial support by the National Research Fund (FNR) Luxembourg.

\end{document}